\documentclass[12pt]{article}
\usepackage{amssymb}
\usepackage{amsmath}
\usepackage[space]{cite}
\usepackage{algorithm}
\usepackage[noend]{algpseudocode}
\usepackage[normalem]{ulem}
\usepackage{graphicx}

\numberwithin{equation}{section}

\newcommand{\be}{\begin{eqnarray}}
\newcommand{\ee}{\end{eqnarray}}
\newcommand{\non}{\nonumber}

\usepackage[usenames,dvipsnames]{color}

\newcommand{\cut}[1]{\ifmmode\text{\textcolor{red}{\sout{\ensuremath{#1}}}}\else\textcolor{red}{\sout{#1}}\fi}

\begin{document}

\begin{titlepage}
\strut\hfill UMTG--312
\vspace{.5in}
\begin{center}

{\LARGE Preparing exact eigenstates of the open XXZ chain\\[0.2in]
on a quantum computer}\\
\vspace{1in}
\large 
John S. Van Dyke\footnote{Department of Physics, Virginia Tech, 
Blacksburg, VA 24061 USA},
Edwin Barnes${}^1$,
Sophia E. Economou${}^1$, \\[0.2in]and
Rafael I. Nepomechie \footnote{Physics Department,
P.O. Box 248046, University of Miami, Coral Gables, FL 33124 USA}
{\let\thefootnote\relax\footnote{{{\tt efbarnes@vt.edu, economou@vt.edu, 
nepomechie@miami.edu, jvandyke@vt.edu}}}}

\end{center}

\vspace{.5in}

\begin{abstract}
The open spin-1/2 XXZ spin chain with diagonal boundary magnetic
fields is the paradigmatic example of a quantum integrable model with
open boundary conditions.  We formulate a quantum algorithm for
preparing Bethe states of this model, corresponding to real solutions
of the Bethe equations.  The algorithm is probabilistic, with a
success probability that decreases with the number of down spins.  For
a Bethe state of $L$ spins with $M$ down spins,
which contains a total of $\binom{L}{M}\, 2^{M}\, M!$ terms, the
algorithm requires $L+M^2+2M$ qubits.
\end{abstract}

\end{titlepage}

\setcounter{footnote}{0}

\section{Introduction}\label{sec:intro}

The existence of {\em exactly solvable} (or {\em quantum integrable}) interacting
many-body quantum models, the first example of which appeared already 
90 years ago \cite{Bethe:1931hc}, is remarkable.
There are infinitely many such models, since they originate
\cite{Faddeev:1996iy} from solutions of the Yang-Baxter equation
\cite{Jimbo:1989qm}, of which there are infinitely many.  These 
models have had, and continue to have, a significant impact in theoretical
physics, ranging from condensed matter physics and statistical 
mechanics to string theory \cite{Batchelor:2007}.

Although much is already known about quantum integrable models, there
is much that remains unknown.  This is due largely to the fact their
exact solutions depend on solutions of corresponding Bethe equations;
and the latter are generally hard to solve.  Hence, despite the fact
that the models are ``exactly solvable,'' significant effort is
generally still necessary to explicitly compute quantities of physical interest.

Quantum computers hold the promise of addressing a variety of heretofore 
intractable problems \cite{Nielsen:2019, Mermin:2007}.
These include quantum simulation of many-body systems, which occur in molecular and
solid-state contexts \cite{Cao:2019, McArdle:2020}. It is natural to
ask whether quantum computers could also help address the
problem of computing quantities of physical interest for quantum
integrable models.  While solving Bethe equations remains an
interesting open challenge \cite{Nepomechie:2021bethe}, an important
recent development is the discovery of an efficient quantum algorithm
for constructing exact eigenstates \cite{VanDyke:2021kvq}.  This algorithm
could potentially be used for explicitly computing correlation functions, which
would otherwise be out of reach.

Integrable models could also impact quantum computing by providing a
testbed for quantum simulators.  While there is a large ongoing effort
to develop near-term algorithms, such as variational quantum
eigensolvers (VQEs) \cite{Cerezo:2021,Bharti:2021}, to solve many-body
problems, it is not clear whether a quantum advantage can be achieved
by VQEs on near-term hardware.  On the other hand, obtaining quantum
advantage for generic simulation problems on fault-tolerant quantum
computers is believed to be enormously costly in terms of quantum
resources \cite{Wecker:2015,Gonthier:2020,vonBurg:2021}.  An
additional benefit of integrable models for early quantum computers
beyond the 
noisy intermediate-scale quantum (NISQ) era \cite{Preskill_2018} 
is that their classically solvable quantities can
be used for validation and verification purposes.  It is therefore
natural to investigate special classes of problems, such as integrable
models, for earlier demonstrations of quantum advantage.  A key first
step is to find quantum algorithms that solve such problems and to
quantify the required resources.

The algorithm in \cite{VanDyke:2021kvq} is for the closed spin-1/2 XXZ spin
chain, which is an anisotropic version \cite{Orbach:1958zz}
of the model solved by Bethe \cite{Bethe:1931hc}, and which is the paradigmatic example of a
quantum integrable model with periodic boundary conditions.  The
extension of quantum integrability to models with {\em open} boundary
conditions is also interesting and nontrivial, see e.g.
\cite{Gaudin:1971zza, Alcaraz:1987uk, Sklyanin:1988yz, 
Ghoshal:1993tm} and related references.

In this paper we formulate a quantum algorithm for 
constructing exact eigenstates of the open spin-1/2 XXZ spin chain with diagonal 
boundary magnetic fields,  which is the paradigmatic example of a
quantum integrable model with open boundary conditions. 
The (ferromagnetic) Hamiltonian ${\cal H}$ for a chain with length $L$ is given by
\be
{\cal H} = -\tfrac{1}{2}\sum_{n=0}^{L-2} \left( 
\sigma^{x}_{n}\, \sigma^{x}_{n+1} + \sigma^{y}_{n}\, \sigma^{y}_{n+1} +
\Delta\, \sigma^{z}_{n}\, \sigma^{z}_{n+1}  \right) 
-\tfrac{1}{2}\left( h\, \sigma^{z}_{0} + h'\, \sigma^{z}_{L-1} \right) \,,
\label{Hamiltonian}
\ee
where as usual $\sigma^{x}_{n}\,, \sigma^{y}_{n}\,, \sigma^{z}_{n}$
are Pauli matrices at site $n$.  The Hamiltonian has three parameters:
the anisotropy parameter $\Delta$, and the boundary magnetic fields
$h$ and $h'$, all of which are assumed here to be real.

An interesting feature of this model is that, for the special values of 
the boundary magnetic fields
\be
h'=-h=\frac{1}{2}(q-q^{-1})\,, \qquad \text{where} \quad
\Delta=\frac{1}{2}(q+q^{-1})\,,
\label{QG}
\ee
the model has the quantum group symmetry $U_{q}(su(2))$ \cite{Pasquier:1989kd, Kulish:1991np}. 
Consequently, the spectrum has the 
degeneracies of an isotropic ($su(2)$-invariant) model, even if 
$\Delta \ne 1$.

While there are many similarities between the closed-chain 
and open-chain algorithms, the latter has some new features due to 
the greater complexity of open-chain Bethe states. As in 
\cite{VanDyke:2021kvq}, we restrict here to 
solutions of the Bethe equations that are real, which can be efficiently determined 
classically, and which serve as inputs for the quantum algorithm. The 
quantum algorithm has as outputs exact eigenstates of the 
Hamiltonian (\ref{Hamiltonian}), with a probability that decreases 
with the number of down spins. We expect that these eigenstates could 
be used to explicitly compute the model's correlation functions, beyond what can be done analytically
\cite{Kitanine:2007bi, Kitanine:2008wb}.

The outline of the remainder of this paper is as follows.  In Sec.
\ref{sec:CBA}, we briefly review the model's coordinate Bethe ansatz solution
\cite{Alcaraz:1987uk}, and we recast it in a form that 
is convenient for our algorithm. We present the algorithm in Sec. 
\ref{sec:algorithm}. In Sec. \ref{sec:sim}, we summarize the results of 
our simulations of this algorithm. We conclude in Sec. \ref{sec:end} with 
a discussion of our results. The complete circuit for the simple case  $L=4, 
M=2$ is provided in Appendix \ref{sec:L4M2}.

\section{Coordinate Bethe ansatz}\label{sec:CBA}

We briefly recall here the coordinate Bethe ansatz solution \cite{Alcaraz:1987uk} of the 
model, and recast it in a form that is convenient for our algorithm, 
see (\ref{permutation})-(\ref{Vfactor}) below. 
(The solution for the case without boundary magnetic fields 
was found in \cite{Gaudin:1971zza}, and the algebraic Bethe ansatz 
solution was formulated in \cite{Sklyanin:1988yz}.)
The Bethe states are exact eigenstates $|M \rangle$ of the Hamiltonian 
(\ref{Hamiltonian})
\be
{\cal H} |M \rangle = E |M \rangle \,,
\ee 
which are given by
\be
|M \rangle = \sum_{0 \le x_{0} < x_{1} < \ldots < x_{M-1} \le L-1} 
f(x_{0}, \ldots, x_{M-1}) |x_{0}, \ldots, x_{M-1} 
\rangle \,,
\label{Bethestate}
\ee
where $x_{0}, \ldots, x_{M-1}$ denote the location of the $M$ down spins. 
In other words,
\be
|x_{0}, \ldots, x_{M-1} \rangle = \sigma^{-}_{x_{0}} \ldots 
\sigma^{-}_{x_{M-1}} |0 \ldots 0 \rangle\,,
\ee
where $\sigma^{-}_{n}=\frac{1}{2}(\sigma^{x}_{n}- i \sigma^{y}_{n})$ is the 
spin-lowering operator at site $n$, and $|0 \ldots 0 \rangle$ is the ferromagnetic ground state
(i.e., the reference state with all $L$ spins in the up-state $|0\rangle = {1\choose 0}$.) 
Moreover, the wave function $f(x_{0}, \ldots, x_{M-1})$ is given by
\be
f(x_{0}, \ldots, x_{M-1}) = \sum_{P} \varepsilon_{P}\, A(k_{0}, 
\ldots, k_{M-1}) e^{i \sum_{j=0}^{M-1} k_{j} x_{j}}\,,
\label{wavefunction}
\ee 
where the sum is over all permutations and negations of $k_{0}, 
\ldots, k_{M-1}$, and $\varepsilon_{P}=\pm 1$ changes sign at each such 
mutation. For example, for $M=2$, the wave function is given by
\begin{align}
f(x_{0}, x_{1}) &= A(k_{0}, k_{1}) e^{i(k_{0} x_{0} + k_{1} x_{1})} 
- A(-k_{0}, k_{1}) e^{i(-k_{0} x_{0} + k_{1} x_{1})} \non\\
& - A(k_{0}, -k_{1}) e^{i(k_{0} x_{0} - k_{1} x_{1})} 
+ A(-k_{0}, -k_{1}) e^{i(-k_{0} x_{0} - k_{1} x_{1})} \non\\
&- A(k_{1}, k_{0}) e^{i(k_{1} x_{0} + k_{0} x_{1})} 
+ A(-k_{1}, k_{0}) e^{i(-k_{1} x_{0} + k_{0} x_{1})} \non\\
& + A(k_{1}, -k_{0}) e^{i(k_{1} x_{0} - k_{0} x_{1})}
-A(-k_{1}, -k_{0}) e^{i(-k_{1} x_{0} - k_{0} x_{1})} \,.
\end{align}
In general, there are $2^{M}\, M!$ terms for fixed $x_0,\ldots,x_{M-1}$, while the 
wave function for the corresponding closed chain 
has only $M!$ terms, since in the latter case there is a sum only over the 
permutations.

The coefficients $A(k_{0}, \ldots, k_{M-1})$ are given by
\be
A(k_{0}, \ldots, k_{M-1}) = \prod_{j=0}^{M-1} \beta(-k_{j}) 
\prod_{0\le j<l\le M-1} B(-k_{j}, k_{l}) e^{-i k_{l}} \,,
\label{Acoeff}
\ee
where 
\be
B(k,k') = s(k, k')\, s(k', k) \,, \qquad s(k, k') = 1 - 2 \Delta 
e^{ik'} + e^{i(k+k')}\,.
\ee
The so-called Bethe roots $k_{0}, \ldots, k_{M-1}$ satisfy the Bethe equations
\be
\frac{\alpha(k_{j})\, \beta(k_{j})}{\alpha(-k_{j})\, \beta(-k_{j})}
= \prod_{l=0; l\ne j}^{M-1} \frac{B(-k_{j}, k_{l})}{B(k_{j}, k_{l})} 
\,, \qquad j = 0, \ldots, M-1 \,,
\label{BE}
\ee
where
\begin{align}
\alpha(k) &= 1 + (h-\Delta)e^{-i k} \,, \label{alpha} \\
\beta(k) &= \left[1 + (h'-\Delta)e^{-i k} \right] e^{i(L+1) k}\,.  
\label{beta} 
\end{align}
The eigenvalues are given by
\be
E(\{k_j\}) = -\tfrac{1}{2}\left[(L-1)\Delta + h + h' \right] +
2\sum_{j=0}^{M-1} (\Delta-\cos(k_{j})) \,.
\label{energy}
\ee

The identity
\be
\frac{s(k',k)}{s(k,k')} = e^{i \Theta(k,k')} \,,
\ee
where $k, k', \Delta$ are real and $\Theta(k, k')$ is defined by
\be
\Theta(k, k') = 2 \arctan\left[\frac{\Delta \sin(\frac{k-k'}{2})}
{\Delta \cos(\frac{k-k'}{2}) - \cos(\frac{k+k'}{2})}\right] \,,
\label{Theta}
\ee
implies that the ratio of $B$-functions appearing in the Bethe 
equations is given by
\be
\frac{B(-k, k')}{B(k, k')} = e^{i \left[\Theta(k,k') + \Theta(k,-k') 
\right]} \,.
\label{logB}
\ee

We observe the further identities
\be
\frac{\alpha(k)}{\alpha(-k)} = e^{i \Phi(k, h)}\,, \qquad
\frac{\beta(k)}{\beta(-k)} = e^{i \Phi(k, h')}\, e^{i 2 (L+1) k}\,, 
\label{logab}
\ee
where $k, h$ are real and the function $\Phi(k, h)$ is defined by
\be
\Phi(k, h) = -2  \arctan\left[\frac{(h-\Delta) \sin(k)}
{1+(h-\Delta)\cos(k)}\right] \,.
\label{Psi}
\ee

Taking the logarithm of the Bethe equations (\ref{BE}) using 
(\ref{logB}) and (\ref{logab}), we obtain
(assuming that all Bethe roots $k_{0}, \ldots, k_{M-1}$ are real)
\be
Z(k_{j}; \{ k_{l} \}) = 2\pi J_{j} \,, \qquad j=0, \ldots, M-1 \,,
\label{logBE}
\ee
where $Z(k; \{ k_{l} \})$ is the so-called counting function
\be
Z(k; \{ k_{l} \}) = 2(L+1)k + \Phi(k,h) + \Phi(k,h') + \Theta(k,-k) -
\sum_{l=0}^{M-1} 
\left[\Theta(k, k_{l}) + \Theta(k, -k_{l}) \right] \,,
\label{counting}
\ee
and $J_{j}$ are distinct integers satisfying
$\{ J_{0}, \ldots, J_{M-1}\} \subset \{1, \ldots, L\}$. In contrast with the 
corresponding closed chain, the counting function (\ref{counting}) 
involves both $\Theta(k, k_{l})$ and $\Theta(k, -k_{l})$, and includes 
terms that depend on the boundary magnetic fields $h$ and $h'$.

In our algorithm for constructing the Bethe states on a quantum 
computer, we compute the Bethe roots 
classically by solving (\ref{logBE}) numerically by iteration. Namely, starting from 
$k_{j}^{(0)} = J_{j}$, we solve  
\be
Z(k_{j}^{(n+1)}; \{ k_{l}^{(n)} \}) = 2\pi J_{j} \,, \qquad j=0, 
\ldots, M-1 \,, \qquad n = 0, 1, \ldots \,,
\label{BEiter}
\ee
which converges rapidly. In this work we consider 
examples with $M \le \lfloor \frac{L}{2} \rfloor$.

We observe that the coefficients (\ref{Acoeff}) satisfy
\be
\frac{A(k_0, \ldots, k_j, k_l, \ldots, k_{M-1})}
{A(k_0, \ldots, k_l, k_j,  \ldots,  k_{M-1})} = e^{i \Theta(k_j, 
k_l)}\,,
\label{permutation}
\ee
as well as 
\be
\frac{A(k_0, \ldots, -k_j, \ldots, k_{M-1})}
{A(k_0, \ldots, k_j, \ldots, k_{M-1})} = e^{i k_j (2L+2)}\, e^{i 
\Phi(k_j, h)}\, 
V(k_j; k_{j+1}, \ldots, k_{M-1}) \,,
\label{negation}
\ee
where
\be
V(k_j; k_{j+1}, \ldots, k_{M-1}) = 
e^{i\sum_{l=j+1}^{M-1} \left[ \Theta(-k_j, k_l) + \Theta(k_l, k_j) 
\right]} \,.
\label{Vfactor}
\ee
While the relation (\ref{permutation}) is true also for the 
corresponding closed chain, the relation (\ref{negation}) is a new 
feature of the open chain. Our algorithm for constructing Bethe 
states is based on the relations (\ref{permutation}) - 
(\ref{Vfactor}).

\section{Algorithm}\label{sec:algorithm}

Our algorithm for preparing the $L$-qubit Bethe state
(\ref{Bethestate}) requires a minimum of $L + M^{2} + 2M$ qubits
(hence, $M^{2} + 2M$ ancillas), which are allocated as follows:
\begin{itemize}
	\item $L$ ``system'' qubits (designated by $s$), whose state 
	becomes the Bethe state (\ref{Bethestate}) on successful completion of the 
	algorithm 
	\item $M(M+1)$ ``permutation-label'' qubits (designated by $p$), which are used to implement 
	the sum over all possible permutations and negations in the wave function 
	(\ref{wavefunction}), and to apply the phases in (\ref{permutation})-(\ref{Vfactor}) 
	and the signs $\varepsilon_{P}$ in (\ref{wavefunction})
	\item $M$ ``faucet'' qubits (designated by $f$), which are used 
	to apply the phase $e^{i \sum_{j=0}^{M-1} k_{j} x_{j}}$ in the 
	wave function (\ref{wavefunction}). 
\end{itemize}
In our implementation of the algorithm, we make use of two additional 
``work'' qubits, in order to reduce the gate count.

The algorithm consists of the steps outlined in Algorithm \ref{alg:Bethe}, 
and which are described in more detail below.

\begin{algorithm}[h]
	\caption{Preparation of the Bethe state (\ref{Bethestate})}\label{alg:Bethe}
	\begin{algorithmic}[1]
		\State Determine the real Bethe roots $\{ k_{0}, \ldots, k_{M-1}\}$ by classically 
		solving (\ref{BEiter}) \label{alg:classical}
		\State Prepare the ``system'' qubits in the Dicke state 
		$\sum_{0 \le x_{0} < \ldots < x_{M-1} \le L-1}  
		|x_{0}, \ldots, x_{M-1} \rangle_{s}$ \label{alg:Dicke}
		\State Prepare the ``permutation-label'' qubits in a superposition state 
		representing all possible permutations and negations; 
		apply the phases in (\ref{permutation})-(\ref{Vfactor})
		and the corresponding signs 
		$\varepsilon_{P}$ in (\ref{wavefunction}) \label{alg:permutation}
		\State Apply the phase $e^{i \sum_{j=0}^{M-1} k_{j} x_{j}}$ 
		in (\ref{wavefunction})
		using the ``faucet'' method \label{alg:faucet}
		\State Reverse the operations on the ancillas (without 
		phases) \label{alg:reverse}
		\State Measure the ``permutation-label'' qubits, with success 
		on $|00\cdots0\rangle_{p}$ \label{alg:measure}
	\end{algorithmic}
\end{algorithm}

\paragraph{Step \ref{alg:classical}}

As already discussed in Sec.  \ref{sec:CBA}, given real values of the
parameters $\Delta$, $h$ and $h'$ and a set of $M$ distinct integers $\{
J_{0}, \ldots, J_{M-1}\} \subset \{1, \ldots, L\}$, the first step
of the algorithm is to determine the corresponding
set of real Bethe roots $\{ k_{0}, \ldots, k_{M-1}\}$ by classically
solving (\ref{BEiter}).  The corresponding energy eigenvalue is given
by (\ref{energy}).

\paragraph{Step \ref{alg:Dicke}}

This step of the algorithm is to prepare the ``system'' qubits
in the so-called Dicke state 
\be
|D_{L,M}\rangle_{s} = \sum_{0 \le x_{0} < \ldots < x_{M-1} \le L-1} |x_{0}, \ldots, 
x_{M-1} \rangle_{s} \,, \non
\ee
which is the same as the Bethe state (\ref{Bethestate})  
but with all the coefficients $f(x_{0}, \ldots, x_{M-1})$ set to 1. This state is an equal-weight 
superposition of the ${L \choose M}$ $L$-qubit states with $M$ 1's 
and $L-M$ 0's. For example,
\be
|D_{3,1}\rangle = \frac{1}{\sqrt{3}}\left( |001\rangle + |010\rangle 
+ |100\rangle \right) \,. \non
\ee
A deterministic algorithm for constructing 
this state is already available \cite{Bartschi2019}, 
see \cite{Mukherjee:2020} for an improved gate count.

\paragraph{Step \ref{alg:permutation}}

The goal of this step of the algorithm is to prepare 
the ``permutation-label'' qubits in the state
\be
\frac{1}{\sqrt{2^{M} M!}}\sum_{P} \varepsilon_{P}\, A(k_{0}, \ldots, 
k_{M-1}) |P\rangle_{p} \,.
\non
\ee
As in \cite{VanDyke:2021kvq}, the phases $\varepsilon_{P}\, A(k_{0}, \ldots, 
k_{M-1})$ are kicked back onto the ``system'' qubits, while 
$|P\rangle_{p}$ is used in step \ref{alg:faucet}
to apply controlled gates. However, here  $|P\rangle_{p}$ represents not only 
permutations, but also negations. The ``permutation-label'' register consists of $M$ 
subregisters, each of which consists of $M+1$ qubits, for a total of $M(M+1)$ qubits.
In each subregister, the first $M$ qubits in the 
subregister, which we call the ``hot'' qubits, 
store an integer in the set $\{0, \ldots, M-1\}$  by means of ``one-hot encoding''. 
As an example, for the case $M=3$, the 
possible states of the ``hot'' qubits
are $|0\rangle \equiv |001\rangle$, $|1\rangle \equiv |010\rangle$, 
$|2\rangle \equiv |100\rangle$. 
These integers serve to label the $M$ down spins in a given
eigenstate and are used to efficiently generate the phases arising
from the permutations that appear in (\ref{wavefunction}).  The last
qubit in each subregister is a ``reflection'' qubit, whose state
indicates the presence (when $|1\rangle$) or absence (when
$|0\rangle$) of a negation associated with the given down spin in the
corresponding term of (\ref{wavefunction}).

To obtain an efficient method for state preparation of
(\ref{Bethestate}), we first implement the phases associated with
negations on the various down spins, while the permutation part is
trivial (i.e. represents only the identity permutation).  We then
execute the permutations and corresponding phase gates on the
permutation-label, to fully produce all terms in (\ref{Bethestate}).
Starting with the ``permutation-label'' qubits in the initial state
$|00\ldots\rangle_{p}$, we encode the identity permutation by storing the 
values $0, \ldots, M-1$ in ascending order (from right to left).  
We next apply Hadamard gates to the reflection qubit in each sublabel
to generate the superposition of having reflected/not reflected each
down spin label. Again for the case $M=3$, the ``permutation-label''
qubits are in the state
\be
\left(H|0\rangle|100\rangle \right)\left(H|0\rangle|010\rangle \right)
\left(H|0\rangle|001\rangle \right) \,.
\label{idperm}
\ee

We next apply the phases in (\ref{negation}). The factor 
$e^{i k_j (2L+2)}\, e^{i \Phi(k_j, h)}$ is simply implemented by applying
to the $j^{th}$ ``reflection'' qubit
a phase gate with angle $k_j (2L+2) +  \Phi(k_j, h) + \pi$; 
the additional phase $\pi$ implements the 
corresponding sign $\varepsilon_{P}$ in (\ref{wavefunction}). 
The factor $V(k_j; k_{j+1}, \ldots, k_{M-1})$ is implemented via an
iterative process.  Working backwards from down spin $j=M-2$ to $j=0$,
the exponential factors for $l>j$ in (\ref{Vfactor}) can be
implemented using a phase gate acting on the reflection qubit of spin
$j$.  In principle, one ought to control on the state of the
reflection qubit of spin $l$, since $k_l$ could be negative.  However,
owing to the identity $\Theta(-k_j,k_l) = \Theta(-k_l,k_j) $ for real
$k_i$ and $\Delta$, one has that
\be
\Theta(-k_j,-k_l) + \Theta(-k_l,k_j) = \Theta(-k_j,k_l) 
+\Theta(k_l,k_j)\,,
\ee
and so the factor ultimately appearing in (\ref{Vfactor}) is the
same regardless of the sign of $k_l$.  Thus, the summation in
(\ref{Vfactor}) can be performed classically and a single phase gate
can be used to produce $V(k_j; k_{j+1}, \ldots, k_{M-1})$ for fixed
$j$.  In fact, the corresponding phase can be added to others
appearing in (\ref{negation}), so that all these can be implemented
simultaneously (along with the sign $\varepsilon_P$).

Starting from the state representing the identity permutation
such as (\ref{idperm}), we now generate a state representing a superposition
of the $M!$ permutations by iteratively swapping subregisters, as
explained in \cite{VanDyke:2021kvq}.  To implement the phase in
(\ref{permutation}), we apply phase gates with angles $\Theta(
\epsilon_{j} k_{j}\,, \epsilon_{l} k_{l}) + \pi$ after the
subregisters $j$ and $l$ have been swapped; these
gates are controlled in part by ``reflection qubits'' to ensure the
proper signs $\epsilon = \pm 1$.  The additional phase $\pi$
implements the corresponding sign $\varepsilon_{P}$ in
(\ref{wavefunction}).

\paragraph{Step \ref{alg:faucet}}

This step of the algorithm is to apply the phase $e^{i
\sum_{j=0}^{M-1} k_{j} x_{j}}$ in (\ref{wavefunction}).  We use the
faucet method \cite{VanDyke:2021kvq}, which exploits the fact that
each $x_{j}$ is an integer; hence, the phase $e^{i \epsilon_{j} k_{j}
x_{j}}$ can be applied by performing $x_{j}$ repeated applications of
the phase $e^{i \epsilon_{j} k_{j}}$.  All $M$ ``faucet'' qubits are
initially turned on (i.e., placed in the state $|1\rangle$).  We then
loop through the $L$ ``system'' qubits.  At each step, we check
to see if the current system qubit represents a down spin (is in the
state $|1\rangle$).  If so, the next faucet qubit is turned off (set
to $|0\rangle$), as the complete phase $e^{i \epsilon_{j} k_{j}
x_{j}}$ for the given $k_{j}$ value has been produced.  
Moreover, as we loop through the ``system'' qubits, we
also apply controlled phase gates with angle $\epsilon_{j} k_{j}$ on the ``hot''
qubits that are controlled by the $j^{th}$ ``faucet'' qubit (must be
on), and by the $j^{th}$ ``reflection'' qubit (determines the sign
$\epsilon_{j}$).

\paragraph{Step \ref{alg:reverse}}

This step of the algorithm is to reverse the operations on the ancillas, 
except without applying any phases. After this step, the 
``faucet'' qubits (but not necessarily all the 
``permutation-label'' qubits) are in the $|0\rangle$ state; and 
the state $|\Psi\rangle$ of the quantum computer is given by
\be
|\Psi\rangle = \alpha |00 \ldots 0\rangle_{p} |00 \ldots 
0\rangle_{f} |M \rangle_{s} + \ldots \,,
\label{result}
\ee
where $|M \rangle_{s}$ is the normalized Bethe state 
(\ref{Bethestate}),  $\alpha$ is some complex number, and the 
ellipsis denotes additional terms that are
orthogonal to the first one. 

\paragraph{Step \ref{alg:measure}}

The final step of the algorithm is to measure the ``permutation-label'' qubits.
It follows from (\ref{result}) that the ``system'' qubits are in the 
target normalized Bethe state $|M \rangle_{s}$ 
on the outcome $|00 \ldots 0\rangle_{p}$, with success probability $|\alpha|^{2}$.

\section{Simulations}\label{sec:sim}

We have implemented Algorithm \ref{alg:Bethe} using Qiskit, and we have 
executed the quantum circuit on the Statevector Simulator for values of $M$ up to 
$M=3$, and for values of $L$ up to $L=8$. We have verified that the 
state obtained by projecting the final state (\ref{result}) to the subspace with all 
ancillas in the state $|0\rangle$ is indeed an exact eigenstate of 
the Hamiltonian (\ref{Hamiltonian}), with eigenvalue (\ref{energy}).

\begin{figure}
\includegraphics[scale=1.0]{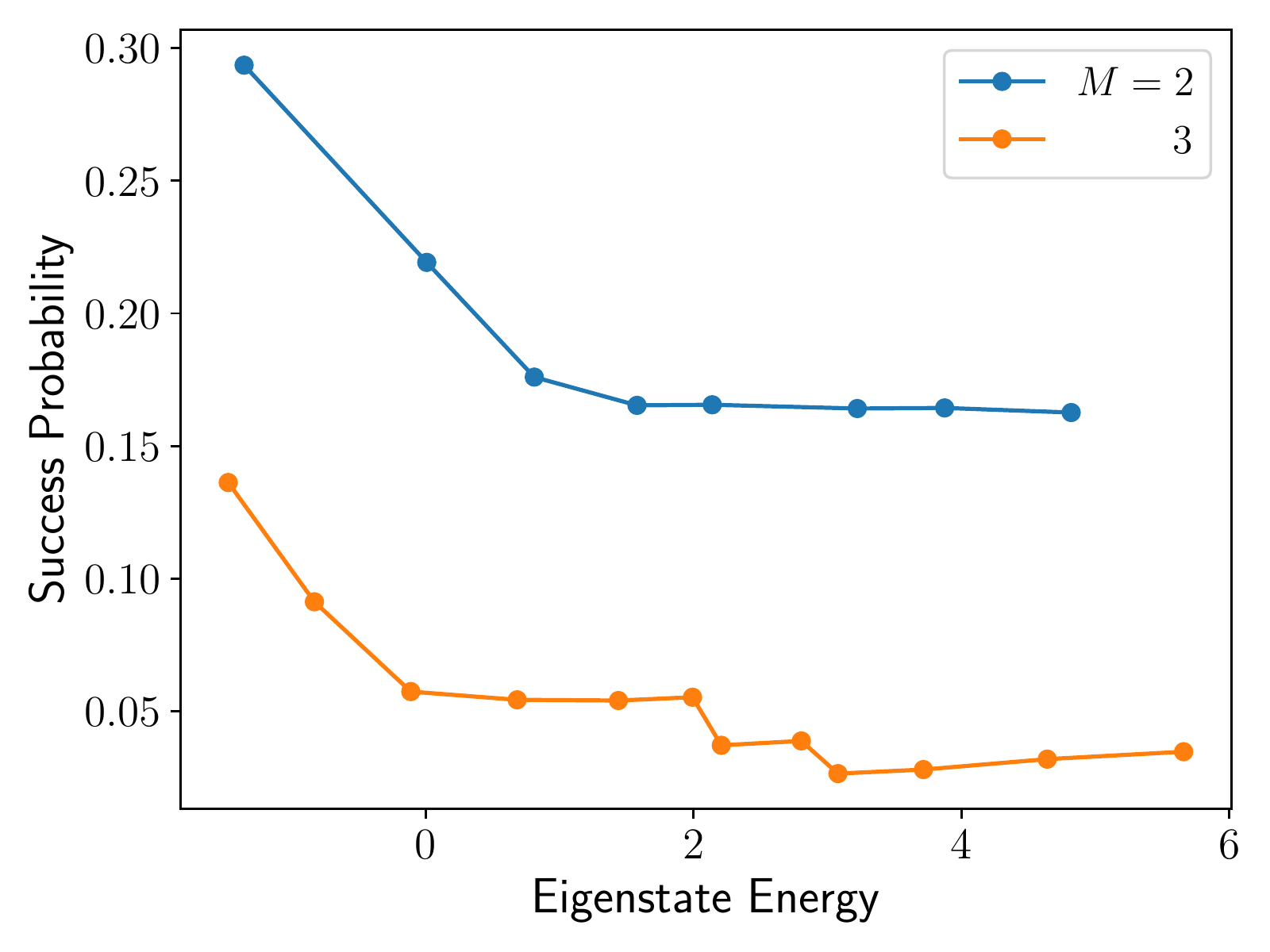}
\caption{Success probability of the open chain Bethe ansatz state
preparation algorithm as a function of eigenstate energy, for $M=2,3$
and $L=6$.  Other parameters are $\Delta = 0.5$, $h=0.1$,
$h'=0.3$.}\label{fig:successprobvaryM}
\end{figure}

\begin{figure}
\includegraphics[scale=1.0]{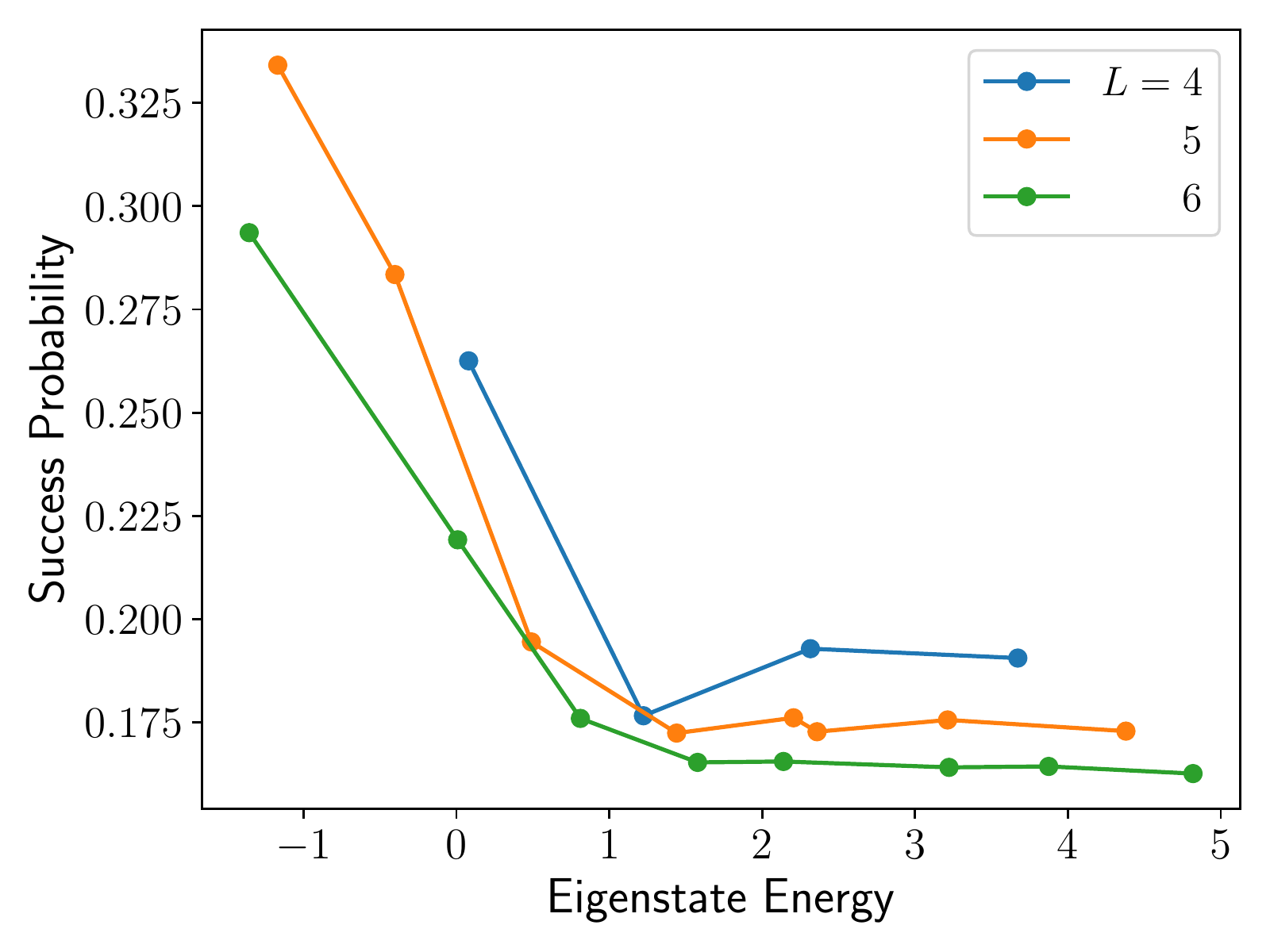}
\caption{Success probability of the open chain Bethe ansatz state
preparation algorithm as a function of eigenstate energy, for
$L=4,5,6$ and $M=2$.  Other parameters are $\Delta = 0.5$, $h=0.1$,
$h'=0.3$.}\label{fig:successprobvaryL}
\end{figure}

Figure \ref{fig:successprobvaryM} shows the success probability
$|\alpha|^{2}$ of the algorithm as a function of eigenstate energy for
$M=2,3$ down spins and $L=6$ sites.  As in the closed chain, there is
a clear decrease of success probability with $M$, as well as with
energy. Figure \ref{fig:successprobvaryL} illustrates the success probability for
fixed $M=2$ and varying $L=4,5,6$.  This also agrees with findings
from the closed chain, in that the minimum success probability does
not vary strongly with $L$.  This suggests that at reasonably small
$M$ one can maintain a high success probability for large $L$ systems,
beyond what is classically tractable with numerical methods.
Furthermore, amplitude amplification can be straightforwardly applied
to enhance the success probability for the open chain, as was
demonstrated explicitly for the closed case \cite{VanDyke:2021kvq}.

\section{Discussion}\label{sec:end}

We have formulated a probabilistic algorithm for preparing exact
eigenstates of the Hamiltonian (\ref{Hamiltonian}).  Due to the
significant depth of the circuit, its implementation on existing and near-term
hardware may not be feasible.  However, we expect that a
fault-tolerant quantum computer with a few hundred qubits will
outperform any classical computer on this task, given that the
algorithm requires only $L+M^2+2M$ qubits, while the Bethe eigenstate (\ref{Bethestate}) 
is given by a sum of $\binom{L}{M}\, 2^{M}\, M!$ terms.
This algorithm is a generalization of the one 
for the corresponding closed chain \cite{VanDyke:2021kvq}, which 
requires $L+M^2+M$ qubits. The additional $M$ ancilla qubits in the open chain case are the reflection qubits that are used to keep track of the negations appearing in (\ref{wavefunction}).

We expect that it will be possible to use the eigenstates furnished by
this algorithm to explicitly compute ground-state correlation functions.  Indeed,
at least for some range of parameter values, the ground state of the
antiferromagnetic Hamiltonian (i.e., $-{\cal H}$, with ${\cal H}$
given by (\ref{Hamiltonian})) is described by $M=L/2$ real Bethe
roots, and can therefore be obtained using this algorithm. Moreover, 
standard routines exist for evaluating expectation values of 
products of Pauli operators on a quantum computer.

Of course, not all eigenstates of this Hamiltonian can be described by
real Bethe roots.  It remains an interesting challenge to prepare
generic eigenstates, which would require solving two main problems:
(1) finding an efficient algorithm for determining the (complex) Bethe
roots; and (2) generalizing the state-preparation algorithm to the
case of complex Bethe roots.  Indeed, for the former problem,
neither a classical nor a quantum algorithm is available; even for the
simpler case of the closed XXX chain with periodic boundary
conditions, the complete set of Bethe roots is known for at most
$L=14$ sites \cite{Hao:2013jqa}.  Regarding the latter problem: for
complex Bethe roots, the factors in (\ref{permutation}),
(\ref{negation}) as well as $e^{i \sum_{j=0}^{M-1} k_{j} x_{j}}$ are no
longer phases, and therefore those factors can no longer be simply
implemented using controlled phase gates. Overcoming this problem 
would have the additional benefit of allowing the treatment of the 
$U_{q}(su(2))$-invariant case (\ref{QG}) with $q$ a root of unity.
In that case the boundary magnetic fields $h, h'$ are no longer real,
and therefore the Hamiltonian (\ref{Hamiltonian}) is no 
longer Hermitian. Nevertheless, its spectrum is real, and can be related to the 
spectra of $c<1$ minimal conformal field theories \cite{Alcaraz:1987uk, Pasquier:1989kd}.

We expect that the algorithm presented here for preparing Bethe states
of the model (\ref{Hamiltonian}) can be extended to other quantum
integrable models with open boundary conditions that have been solved 
by coordinate Bethe ansatz (e.g. \cite{Asakawa:1996, Fireman_2002}). 
It would also be interesting to
formulate such algorithms that are based instead on algebraic Bethe 
ans\"atze, with which models have also been solved 
(see e.g. \cite{Sklyanin:1988yz, Guan:2000, Li_2007, Belliard_2009, 
Gerrard:2020} and references therein). Although a 
deterministic approach for constructing algebraic Bethe ansatz states
seems difficult \cite{Nepomechie:2021bethe}, a probabilistic approach based on the linear combination of unitaries method \cite{Childs:2012,Berry:2015}
 --  the strategy employed here and in \cite{VanDyke:2021kvq} -- may 
be more feasible.

\section*{Acknowledgments}

E.B. acknowledges support from the Department of Energy, Award No. DE-SC0019199.
S.E.E. acknowledges the DOE Office of Science, National Quantum Information Science Research Centers,
Co-design Center for Quantum Advantage (C2QA), contract number DE-SC0012704.

\appendix

\section{Circuit for $L=4, M=2$}\label{sec:L4M2}

\includegraphics[scale=0.50]{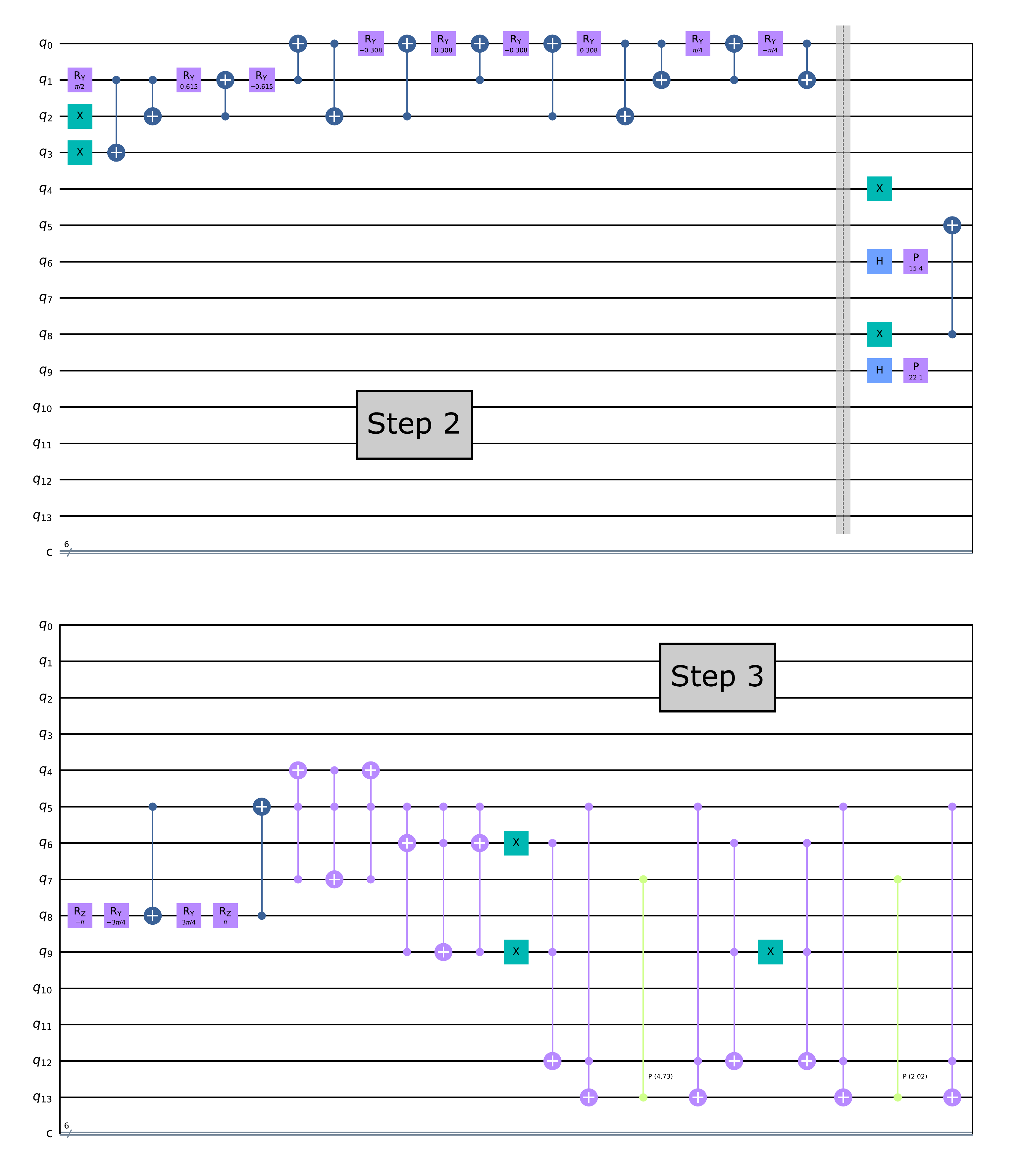}
\newpage
\includegraphics[scale=0.50]{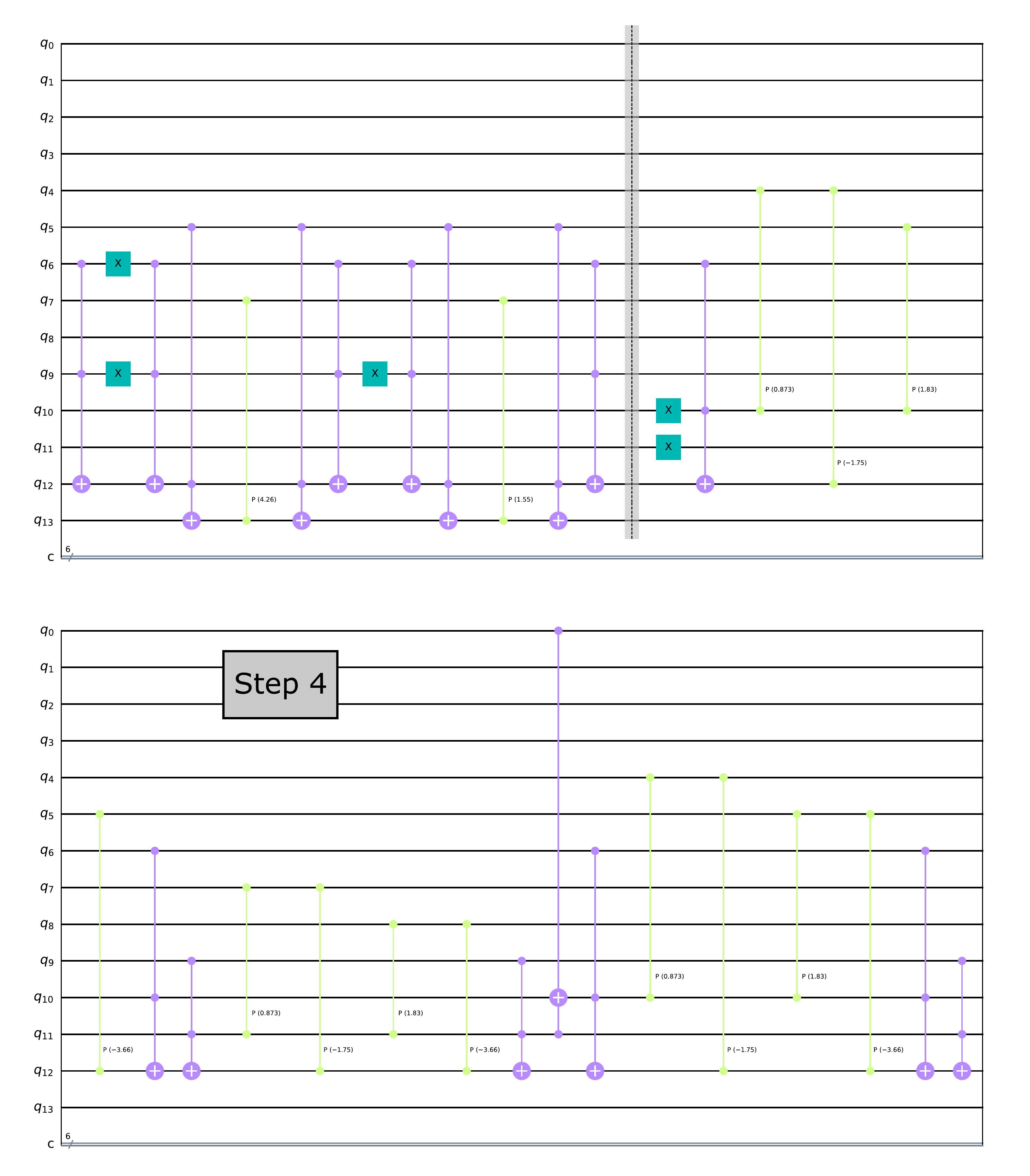}
\newpage
\includegraphics[scale=0.38]{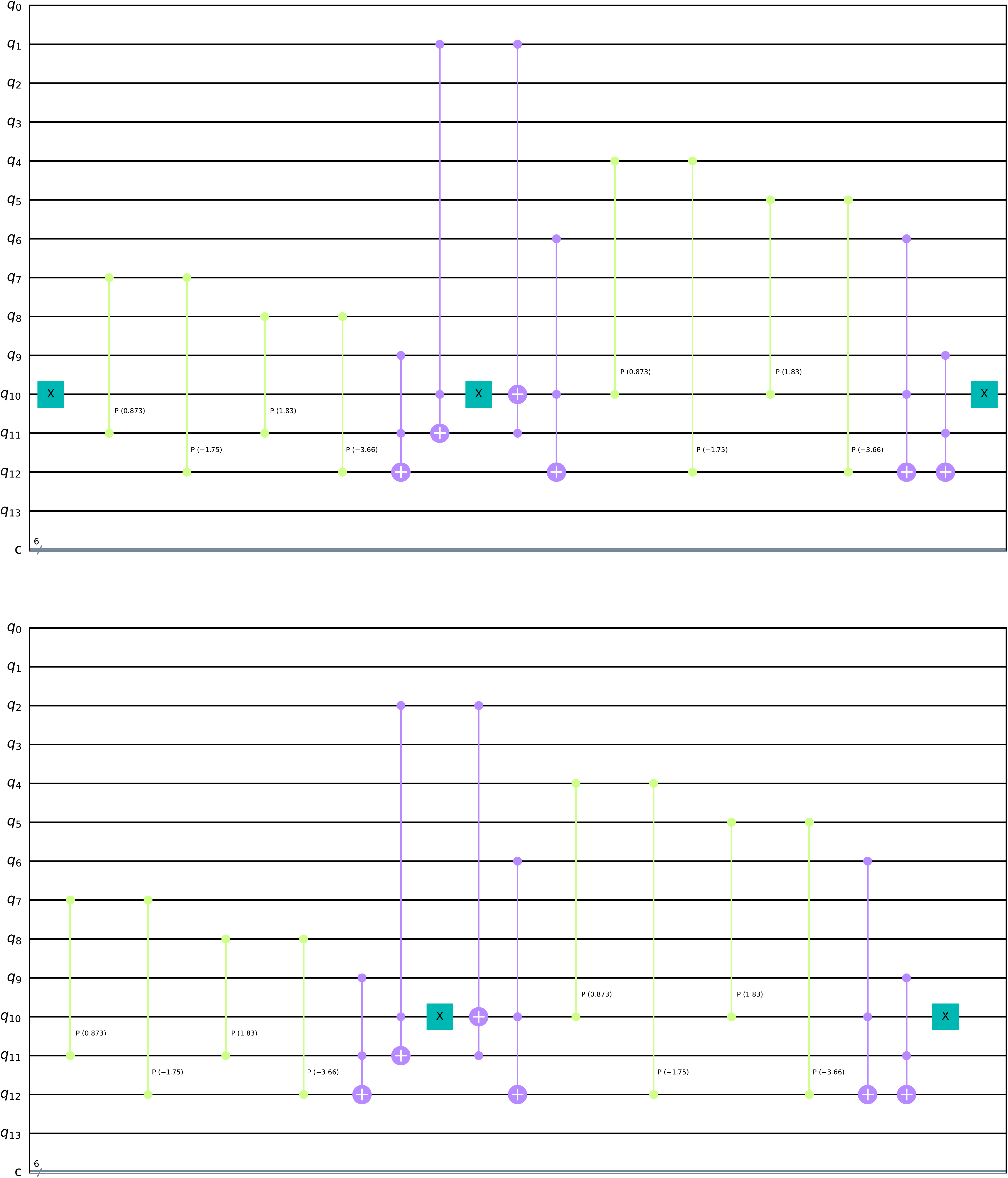}
\newpage

\begin{figure}[h!]
\includegraphics[scale=0.50]{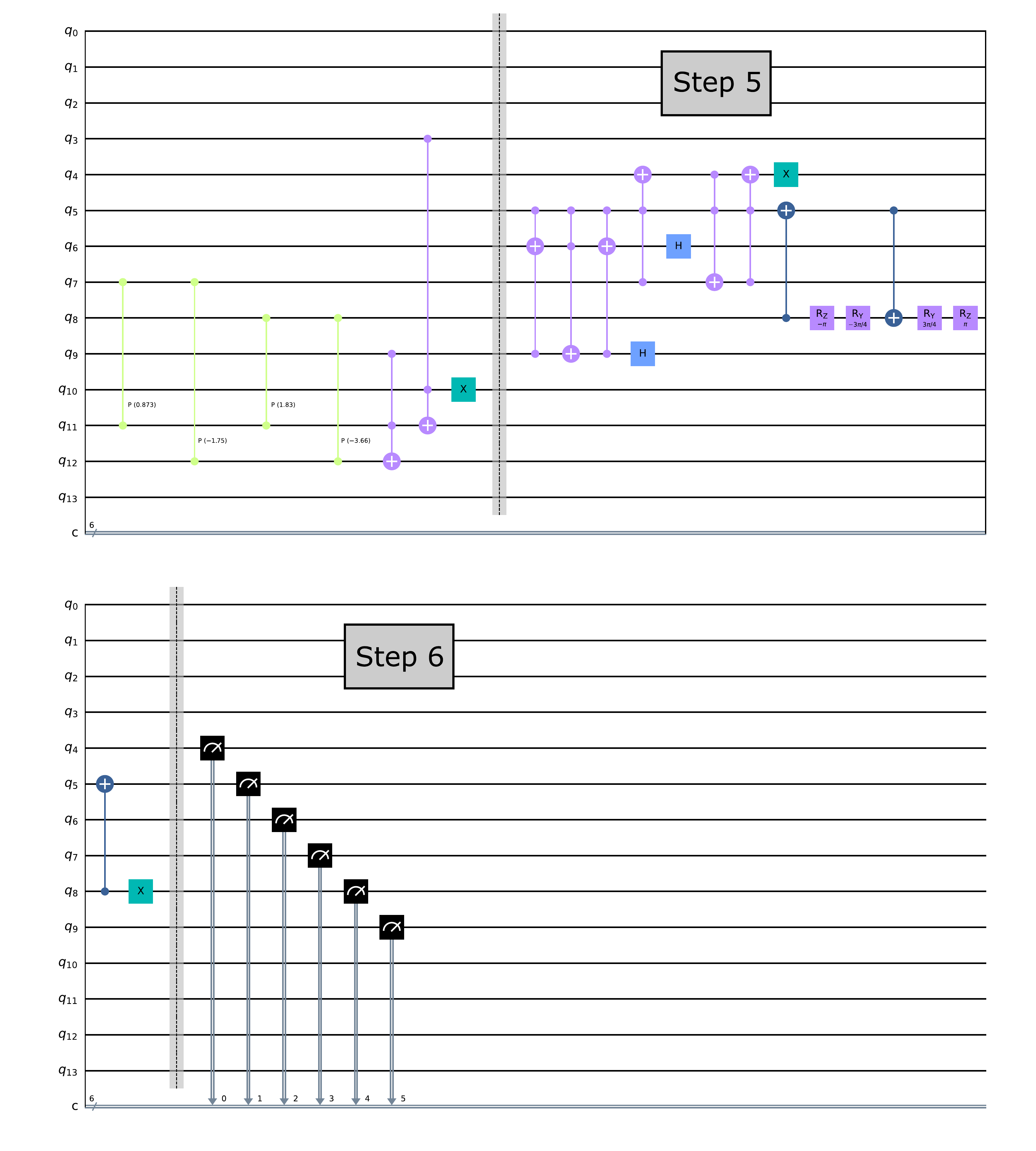}
\caption{Quantum circuit to prepare an eigenstate of the open XXZ
chain for $L=4$ and $M=2$.  Other parameters are $\Delta = 0.5$,
$h=0.1$, $h'=0.3$.  The eigenstate is specified by the Bethe roots
$k_1 = 0.8725655419522633$ and $k_2 =
1.8281634948690795$. 
The various steps of the algorithm, which are separated by barriers, are
labeled; these steps are described in the corresponding parts of Sec. \ref{sec:algorithm}.}
\label{fig:openchaincircuit}
\end{figure}


\begin{thebibliography}{10}

\bibitem{Bethe:1931hc}
H.~Bethe, ``{On the theory of metals. 1. Eigenvalues and eigenfunctions for the
  linear atomic chain},''
\href{http://dx.doi.org/10.1007/BF01341708}{{\em Z. Phys.} {\bfseries 71}
  (1931) 205--226}.

\bibitem{Faddeev:1996iy}
L.~D. Faddeev, ``{How algebraic Bethe ansatz works for integrable models},'' in
  {\em Sym\'etries Quantiques (Les Houches Summer School Proceedings vol 64)},
  A.~Connes, K.~Gawedzki, and J.~Zinn-Justin, eds., pp.~149--219.
\newblock North Holland, 1998.
\newblock
\href{http://arxiv.org/abs/hep-th/9605187}{{\ttfamily arXiv:hep-th/9605187
  [hep-th]}}.
\newblock

\bibitem{Jimbo:1989qm}
M.~Jimbo, ``{Introduction to the {Yang-Baxter} Equation},''
  \href{http://dx.doi.org/10.1142/S0217751X89001503}{{\em Int. J. Mod. Phys. A}
  {\bfseries 4} (1989) 3759--3777}.

\bibitem{Batchelor:2007}
M.~T. Batchelor, ``{The Bethe ansatz after 75 years},''
  \href{http://dx.doi.org/10.1063/1.2709557}{{\em Phys. Today} {\bfseries 60}
  (2007) 36}.

\bibitem{Nielsen:2019}
M.~A. Nielsen and I.~L. Chuang, {\em {Quantum computation and quantum
  information}}.
\newblock Cambridge University Press, 2019.

\bibitem{Mermin:2007}
N.~D. Mermin, {\em {Quantum computer science, an introduction}}.
\newblock Cambridge University Press, 2007.

\bibitem{Cao:2019}
Y.~Cao, J.~Romero, J.~P. Olson, M.~Degroote, P.~D. Johnson, M.~Kieferov{\'a},
  I.~D. Kivlichan, T.~Menke, B.~Peropadre, N.~P.~D. Sawaya, S.~Sim, L.~Veis,
  and A.~Aspuru-Guzik, ``Quantum {Chemistry} in the {Age} of {Quantum}
  {Computing},'' \href{http://dx.doi.org/10.1021/acs.chemrev.8b00803}{{\em
  Chemical Reviews} {\bfseries 119} no.~19, (Aug., 2019) 10856--10915},
  \href{http://arxiv.org/abs/1812.09976}{{\ttfamily arXiv:1812.09976
  [quant-ph]}}.

\bibitem{McArdle:2020}
S.~McArdle, S.~Endo, A.~Aspuru-Guzik, S.~C. Benjamin, and X.~Yuan, ``Quantum
  computational chemistry,''
  \href{http://dx.doi.org/10.1103/RevModPhys.92.015003}{{\em Reviews of Modern
  Physics} {\bfseries 92} no.~1, (Mar., 2020) 015003},
  \href{http://arxiv.org/abs/1808.10402}{{\ttfamily arXiv:1808.10402
  [quant-ph]}}.

\bibitem{Nepomechie:2021bethe}
R.~I. Nepomechie, ``{Bethe ansatz on a quantum computer?},'' {\em Quantum
  Information \& Computation} {\bfseries 21} (2021) 255--265,
  \href{http://arxiv.org/abs/2010.01609}{{\ttfamily arXiv:2010.01609
  [quant-ph]}}.

\bibitem{VanDyke:2021kvq}
J.~S. Van~Dyke, G.~S. Barron, N.~J. Mayhall, E.~Barnes, and S.~E. Economou,
  ``{Preparing Bethe Ansatz Eigenstates on a Quantum Computer},'' {\em PRX
  Quantum} {\bfseries 2} (2021) 040329,
  \href{http://arxiv.org/abs/2103.13388}{{\ttfamily arXiv:2103.13388
  [quant-ph]}}.

\bibitem{Cerezo:2021}
M.~Cerezo, A.~Arrasmith, R.~Babbush, S.~C. Benjamin, S.~Endo, K.~Fujii, J.~R.
  McClean, K.~Mitarai, X.~Yuan, L.~Cincio, and P.~J. Coles, ``Variational
  quantum algorithms,''
  \href{http://dx.doi.org/10.1038/s42254-021-00348-9}{{\em Nature Reviews
  Physics} {\bfseries 3} no.~9, (Sept., 2021) 625--644},
  \href{http://arxiv.org/abs/2012.09265}{{\ttfamily arXiv:2012.09265
  [quant-ph]}}.

\bibitem{Bharti:2021}
K.~Bharti, A.~Cervera-Lierta, T.~H. Kyaw, T.~Haug, S.~Alperin-Lea, A.~Anand,
  M.~Degroote, H.~Heimonen, J.~S. Kottmann, T.~Menke, W.-K. Mok, S.~Sim, L.-C.
  Kwek, and A.~Aspuru-Guzik, ``Noisy intermediate-scale quantum ({NISQ})
  algorithms,'' \href{http://arxiv.org/abs/2101.08448}{{\ttfamily
  arXiv:2101.08448 [quant-ph]}}.

\bibitem{Wecker:2015}
D.~Wecker, M.~B. Hastings, and M.~Troyer, ``Progress towards practical quantum
  variational algorithms,''
  \href{http://dx.doi.org/10.1103/PhysRevA.92.042303}{{\em Physical Review A}
  {\bfseries 92} no.~4, (Oct., 2015) 042303},
  \href{http://arxiv.org/abs/1507.08969}{{\ttfamily arXiv:1507.08969
  [quant-ph]}}.

\bibitem{Gonthier:2020}
J.~F. Gonthier, M.~D. Radin, C.~Buda, E.~J. Doskocil, C.~M. Abuan, and
  J.~Romero, ``Identifying challenges towards practical quantum advantage
  through resource estimation: the measurement roadblock in the variational
  quantum eigensolver,'' \href{http://arxiv.org/abs/2012.04001}{{\ttfamily
  arXiv:2012.04001 [quant-ph]}}.

\bibitem{vonBurg:2021}
V.~von Burg, G.~H. Low, T.~H{\"a}ner, D.~S. Steiger, M.~Reiher, M.~Roetteler,
  and M.~Troyer, ``Quantum computing enhanced computational catalysis,''
  \href{http://dx.doi.org/10.1103/PhysRevResearch.3.033055}{{\em Physical
  Review Research} {\bfseries 3} no.~3, (July, 2021) 033055},
  \href{http://arxiv.org/abs/2007.14460}{{\ttfamily arXiv:2007.14460
  [quant-ph]}}.

\bibitem{Preskill_2018}
J.~Preskill, ``{Quantum Computing in the NISQ era and beyond},''
  \href{http://dx.doi.org/10.22331/q-2018-08-06-79}{{\em Quantum} {\bfseries 2}
  (Aug, 2018) 79}, \href{http://arxiv.org/abs/1801.00862}{{\ttfamily
  arXiv:1801.00862 [quant-ph]}}.

\bibitem{Orbach:1958zz}
R.~Orbach, ``{Linear Antiferromagnetic Chain with Anisotropic Coupling},''
  \href{http://dx.doi.org/10.1103/PhysRev.112.309}{{\em Phys. Rev.} {\bfseries
  112} (1958) 309--316}.

\bibitem{Gaudin:1971zza}
M.~Gaudin, ``{Boundary energy of a Bose gas in one dimension},''
  \href{http://dx.doi.org/10.1103/PhysRevA.4.386}{{\em Phys. Rev. A} {\bfseries
  4} (1971) 386--394}.

\bibitem{Alcaraz:1987uk}
F.~C. Alcaraz, M.~N. Barber, M.~T. Batchelor, R.~J. Baxter, and G.~R.~W.
  Quispel, ``{Surface exponents of the quantum XXZ, Ashkin-Teller and Potts
  models},''
\href{http://dx.doi.org/10.1088/0305-4470/20/18/038}{{\em J. Phys.} {\bfseries
  A20} (1987) 6397}.

\bibitem{Sklyanin:1988yz}
E.~K. Sklyanin, ``{Boundary conditions for integrable quantum systems},''
\href{http://dx.doi.org/10.1088/0305-4470/21/10/015}{{\em J. Phys.} {\bfseries
  A21} (1988) 2375}.

\bibitem{Ghoshal:1993tm}
S.~Ghoshal and A.~B. Zamolodchikov, ``{Boundary S matrix and boundary state in
  two-dimensional integrable quantum field theory},''
  \href{http://dx.doi.org/10.1142/S0217751X94001552}{{\em Int. J. Mod. Phys.}
  {\bfseries A9} (1994) 3841--3886},
  \href{http://arxiv.org/abs/hep-th/9306002}{{\ttfamily arXiv:hep-th/9306002
  [hep-th]}}.
[Erratum: Int. J. Mod. Phys.A9,4353 (1994)].

\bibitem{Pasquier:1989kd}
V.~Pasquier and H.~Saleur, ``{Common structures between finite systems and
  conformal field theories through quantum groups},''
\href{http://dx.doi.org/10.1016/0550-3213(90)90122-T}{{\em Nucl. Phys.}
  {\bfseries B330} (1990) 523--556}.

\bibitem{Kulish:1991np}
P.~P. Kulish and E.~K. Sklyanin, ``{The general $U_q(sl(2))$ invariant XXZ
  integrable quantum spin chain},''
{\em J. Phys.} {\bfseries A24} (1991) L435--L439.

\bibitem{Kitanine:2007bi}
N.~Kitanine, K.~K. Kozlowski, J.~M. Maillet, G.~Niccoli, N.~A. Slavnov, and
  V.~Terras, ``{Correlation functions of the open XXZ chain I},''
  \href{http://dx.doi.org/10.1088/1742-5468/2007/10/P10009}{{\em J. Stat.
  Mech.} {\bfseries 0710} (2007) P10009},
  \href{http://arxiv.org/abs/0707.1995}{{\ttfamily arXiv:0707.1995 [hep-th]}}.

\bibitem{Kitanine:2008wb}
N.~Kitanine, K.~K. Kozlowski, J.~M. Maillet, G.~Niccoli, N.~A. Slavnov, and
  V.~Terras, ``{Correlation functions of the open XXZ chain II},''
  \href{http://dx.doi.org/10.1088/1742-5468/2008/07/P07010}{{\em J. Stat.
  Mech.} {\bfseries 0807} (2008) P07010},
  \href{http://arxiv.org/abs/0803.3305}{{\ttfamily arXiv:0803.3305 [hep-th]}}.

\bibitem{Bartschi2019}
A.~B\"artschi and S.~Eidenbenz, ``{Deterministic preparation of Dicke
  states},'' {\em Lecture Notes in Computer Science} (2019) 126--139,
  \href{http://arxiv.org/abs/1904.07358}{{\ttfamily arXiv:1904.07358
  [quant-ph]}}.

\bibitem{Mukherjee:2020}
C.~S. Mukherjee, S.~Maitra, V.~Gaurav, and D.~Roy, ``{On actual preparation of
  Dicke state on a quantum computer},'' {\em IEEE Trans. Quant. Eng.}
  {\bfseries 1} (2020) 3102517,
  \href{http://arxiv.org/abs/2007.01681}{{\ttfamily arXiv:2007.01681
  [quant-ph]}}.

\bibitem{Hao:2013jqa}
W.~Hao, R.~I. Nepomechie, and A.~J. Sommese, ``{Completeness of solutions of
  Bethe's equations},''
  \href{http://dx.doi.org/10.1103/PhysRevE.88.052113}{{\em Phys. Rev. E}
  {\bfseries 88} no.~5, (2013) 052113},
  \href{http://arxiv.org/abs/1308.4645}{{\ttfamily arXiv:1308.4645 [math-ph]}}.

\bibitem{Asakawa:1996}
H.~{Asakawa} and M.~{Suzuki}, ``{Finite-size corrections in the XXZ model and
  the Hubbard model with boundary fields},''
  \href{http://dx.doi.org/10.1088/0305-4470/29/2/004}{{\em Journal of Physics A
  Mathematical General} {\bfseries 29} no.~2, (Jan., 1996) 225--245}.

\bibitem{Fireman_2002}
E.~Fireman, A.~Lima-Santos, and W.~Utiel, ``Bethe ansatz solution for quantum
  spin-1 chains with boundary terms,''
  \href{http://dx.doi.org/10.1016/s0550-3213(02)00027-5}{{\em Nuclear Physics
  B} {\bfseries 626} no.~3, (Apr, 2002) 435--462},
  \href{http://arxiv.org/abs/0110048}{{\ttfamily arXiv:0110048 [nlin]}}.

\bibitem{Guan:2000}
X.-W. Guan, ``Algebraic {Bethe} ansatz for the one-dimensional {Hubbard} model
  with open boundaries,''
  \href{http://dx.doi.org/10.1088/0305-4470/33/30/309}{{\em Journal of Physics
  A: Mathematical and General} {\bfseries 33} no.~30, (Aug., 2000) 5391},
  \href{http://arxiv.org/abs/9908054}{{\ttfamily arXiv:9908054 [cond-mat]}}.

\bibitem{Li_2007}
G.-L. Li and K.-J. Shi, ``The algebraic {Bethe} ansatz for open vertex
  models,'' \href{http://dx.doi.org/10.1088/1742-5468/2007/01/p01018}{{\em
  Journal of Statistical Mechanics: Theory and Experiment} {\bfseries 2007}
  no.~01, (Jan, 2007) P01018--P01018},
  \href{http://arxiv.org/abs/0611127}{{\ttfamily arXiv:0611127 [hep-th]}}.

\bibitem{Belliard_2009}
S.~Belliard and E.~Ragoucy, ``The nested {Bethe} ansatz for ``all'' open spin
  chains with diagonal boundary conditions,''
  \href{http://dx.doi.org/10.1088/1751-8113/42/20/205203}{{\em Journal of
  Physics A: Mathematical and Theoretical} {\bfseries 42} no.~20, (Apr, 2009)
  205203}, \href{http://arxiv.org/abs/0902.0321}{{\ttfamily arXiv:0902.0321
  [math-ph]}}.

\bibitem{Gerrard:2020}
A.~{Gerrard} and V.~{Regelskis}, ``{Nested algebraic Bethe ansatz for
  orthogonal and symplectic open spin chains},''
  \href{http://dx.doi.org/10.1016/j.nuclphysb.2019.114909}{{\em Nuclear Physics
  B} {\bfseries 952} (Mar., 2020) 114909},
  \href{http://arxiv.org/abs/1909.12123}{{\ttfamily arXiv:1909.12123
  [math-ph]}}.

\bibitem{Childs:2012}
A.~M. Childs and N.~Wiebe, ``Hamiltonian simulation using linear combinations
  of unitary operations,'' {\em Quantum Information \& Computation} {\bfseries
  12} (November, 2012) 901, \href{http://arxiv.org/abs/1202.5822}{{\ttfamily
  arXiv:1202.5822 [quant-ph]}}.

\bibitem{Berry:2015}
D.~W. Berry, A.~M. Childs, R.~Cleve, R.~Kothari, and R.~D. Somma, ``Simulating
  {Hamiltonian} {Dynamics} with a {Truncated} {Taylor} {Series},''
  \href{http://dx.doi.org/10.1103/PhysRevLett.114.090502}{{\em Physical Review
  Letters} {\bfseries 114} no.~9, (Mar., 2015) 090502},
  \href{http://arxiv.org/abs/1412.4687}{{\ttfamily arXiv:1412.4687
  [quant-ph]}}.

\end{thebibliography}
\providecommand{\href}[2]{#2}\begingroup\raggedright\endgroup

\end{document}